\documentclass{article}
\usepackage{graphicx}
\usepackage{amsmath}
\usepackage{amsfonts}
\usepackage{amssymb}
\newtheorem{theorem}{Theorem}

\begin{document}

\title{How far does the analogy between causal horizon-induced thermalization with
the standard heat bath situation go? \\{\small Based on a talk given at the 2002 Winter School in Londrina, Parana, Brazil}}
\author{Bert Schroer\\present address: CBPF, Rua Dr. Xavier Sigaud 150, \\22290-180 Rio de Janeiro, Brazil\\email schroer@cbpf.br\\permanent address: Institut f\"{u}r Theoretische Physik\\FU-Berlin, Arnimallee 14, 14195 Berlin, Germany}
\date{January 2002}
\maketitle
\begin{abstract}
After a short presentation of KMS states and modular theory as the unifying
description of thermalizing systems we propose the absence of transverse
vacuum fluctuations in the holographic projections as the principle reason for
an area behavior (the transverse area) of \textit{localization entropy} as
opposed to the volume dependence of ordinary heat bath entropy. Thermalization
through causal localization is not a property of QM but results from the
omnipresent vacuum polarization in QFT and does not require a Gibbs type
ensemble avaraging (coupling to a heat bath).
\end{abstract}

\section{$_{{}}$Posing the problem}

Although thermal aspects which result from causal localization in local
quantum physics permit a unified description with those generated by the
standard coupling to a heat bath, there are some characteristic and important
physical differences.

Whereas the temperature of local quantum subsystem which are causally
protected from the rest of the world (e.g. a Hawking black hole \cite{Hawking}
or one of its recent non gravitational analogs \cite{U}) is determined by the
geometry of that situation, the heat bath temperature can be continuously
varied without making geometric changes (even if there should exist a maximal
Hagedorn temperature as sometimes envisage in string theory). According to
Bekenstein's bold guess \cite{Beken} about an area behavior of a black hole
entropy (related to Hawking quantum aspects by the postulated classical
fundamental Gibbs form of the 2nd fundamental thermodynamic law which made it
more convincing), one should expect that for any successful attempt to define
directly a quantum entropy associated with a causal horizon it should take the
form of an area density as compared to the volume density in the standard heat
bath case (with the area being the two-dimensional edge of a bifurcated
horizon). The existence of the mentioned non gravitational analogs points
towards a still poorly understood fundamental relation between geometry and
thermal aspects of local quantum physics at a place where one would rather
have expected the (elusive) quantum gravity or string theory.

In these lecture notes we will show that there is an important aspect of a
properly defined \textit{algebraic holographic projection} onto the horizon
which sets the stage for an area density, namely the total absence of vacuum
polarization on the horizon in directions transverse to the light ray
direction \cite{S1}\cite{S2}. We explicitly illustrate this
phenomenon\footnote{According to the authors best knowledge this is the only
case in which the quantum mechanical fluctuationless vacuum structure (and an
ensuing transverse Galilei symmetry) appears in the midst of QFT without
having done any nonrelativistic approximation.} in the case of the
Rindler-Unruh \cite{Unruh} wedge situation (for which the linear extension of
the horizon is the lightfront) and argue that it any potential entropy-like
measure for the impurity which results from restricting the vacuum to the
horizon must necessarily follow an area law where the area is that of the edge
of the wedge or its horizon. Without the normalizing use of a (still unknown)
second fundamental law derived in the setting of horizon-caused quantum
thermal behavior one can at best obtain a \textit{relative} area density which
determines the relation for different quantum matter content. Before this we
will briefly sketch the unifying formalism for both kinds of thermal manifestation.

\section{How modular theory of operator algebras unifies thermal aspects.}

Let us briefly indicate how one gets from the standard description of heat
bath coupled Gibbs ensembles to the more general unifying framework. The
correlation functions of a QFT in a quantization box V (in order to obtain a
discrete energy-momentum spectrum) coupled to a heat bath reservoir are
computed with the well-known Gibbs formula
\begin{align}
\omega_{\beta}(A) &  :=\frac{1}{Z_{V}}tre^{-\beta H_{V}}A,\,\,Z_{V}%
=tre^{-\beta H_{V}}\\
&  \curvearrowright\omega_{\beta}(\mathbf{1)}=1\nonumber
\end{align}
which assigns a (normalized) state\footnote{The existence of inequivalent
representations in the presence of infinitly many degrees of freedom and the
structure of local algebras requires to make a distinction between a state (in
the sense of an expectation value) and a state vector which implements this
state (this is not necessary in $B(\mathcal{H})$ algebras of QM which relates
states one to one with unit rays). } on the algebra of bounded operators.$A\in
\mathcal{A}=B(\mathcal{H}).$ The Gibbs formula is meaningful as long as the
partition function $Z$ exists (which requires a discrete Hamiltonian spectrum
bounded below and with finite degeneracy, i.e. ''boxed'' systems). The
difference to the vacuum situation becomes more visible on the level of the
operator formalism which may be obtained from the state $\omega_{\beta}%
(\cdot)$ on $\mathcal{A}$ ($\simeq$ set of correlation functions) by the
canonical GNS (Gelfand, Neumark and Segal) construction \cite{Haag}. Using the
special property of density matrix states, one may implement the abstract GNS
construction on a Hilbert space $\mathcal{H}_{HS}$ whose vectors are
Hilbert-Schmidt operators $\kappa$ i.e. $tr\kappa^{\ast}\kappa<\infty$%
\begin{align}
\mathcal{H}_{HS} &  =\left\{  \psi_{\kappa}\text{ }|\text{ }(\psi_{\kappa
},\psi_{\kappa^{\prime}})\equiv tr\kappa^{\ast}\kappa^{\prime}\right\}
\label{left}\\
&  \pi(A)\psi_{\kappa}\equiv\psi_{A\kappa}\in\mathcal{H}_{HS}\nonumber
\end{align}
where $\pi(\cdot)$ denotes the representation of the algebra on $\mathcal{H}%
_{HS}.$ The HS Hilbert space is isomorphic to the tensor product of the
original Hilbert space $\mathcal{H}_{HS}\simeq\mathcal{H}\otimes\mathcal{H}$
since the linear combinations of ``dyads'' $\left|  \psi\right\rangle
\left\langle \varphi\right|  $ from the tensor product upon closure in
$\mathcal{H}_{HS}$ generate the HS Hilbert space. This ``doubling'' entails
that besides the left action (\ref{left}) of the full algebra of bounded
operators $B(\mathcal{H})$ on $\mathcal{H}_{HS}$ there is a right action which
in the HS description reads $\psi_{\kappa}\rightarrow\psi_{\kappa A}$
\cite{Haag}$.$ In order to distinguish between the left and right
representation and to maintain the naturality of composition (representation)
laws, one defines the right representation as a conjugate (antilinear)
representation
\begin{align}
\pi_{l}(A)\psi_{\kappa} &  =\pi(A)\psi_{\kappa}=\psi_{A\kappa}\label{doubling}%
\\
\pi_{r}(A)\psi_{\kappa} &  =\psi_{\kappa A^{\ast}}\nonumber
\end{align}
It is obvious that any right action commutes with any left action i.e.
$\pi_{r}(\mathcal{A})\subseteq\pi(\mathcal{A})^{\prime}$ (where the dash
denotes the von Neumann commutant of $\pi(\mathcal{A})$ in $\mathcal{H}_{HS})$
and in this particular case Haag, Hugenholtz and Winnink in their seminal
paper had no problem to prove that in fact equality holds \cite{HHW}. In
$\mathcal{H}_{HS}\simeq\mathcal{H}\otimes\mathcal{H}$ there are many more
operators than in $\pi(\mathcal{A}),$ e.g. the anti-unitary ``flip'' $J$
\begin{align}
J\psi_{\kappa} &  :=\psi_{\kappa\ast},\text{ }J^{2}=1\\
&  J\pi(A)J=\pi_{r}(A)\nonumber
\end{align}
which in the tensor product representation would simply interchange the bra
and ket in a dyad.

Using now the fact that the Hilbert-Schmidt operator $\kappa_{0}\equiv$
$\rho^{\frac{1}{2}}$associated with the nondegenerate (no zero eigenvalue)
Gibbs density matrix
\begin{equation}
\omega_{\beta}(A)=\left(  \psi_{\kappa_{0}},\pi(A)\psi_{\kappa_{0}}\right)
=tr\kappa_{0}A\kappa_{0}%
\end{equation}
is cyclic and separating with respect to the action $\pi(\mathcal{A})$ of the
algebra (i.e. sufficiently entangled in $\mathcal{H}\otimes\mathcal{H}$ so
that the application of this subalgebra permits to approximate any vector in
$\mathcal{H}\otimes\mathcal{H}$ and that it is not possible to annihilate the
entangled state with a nonzero operator from $\pi(\mathcal{A})$), one checks
the validity of the relation (mainly an exercise in the correct application of
definitions)
\begin{align}
S\pi(A)\psi_{\kappa_{0}}  &  =\pi(A)^{\ast}\psi_{\kappa_{0}},\,A\in
\mathcal{A}\\
where\,\,S  &  :=J\pi(\rho^{\frac{1}{2}})\pi_{r}(\rho^{-\frac{1}{2}%
})\curvearrowright S^{2}\subset1\nonumber
\end{align}
where the last relation is a notation for the fact that the
unbounded\footnote{Since $\pi_{r}(\rho^{-\frac{1}{2}})$ is unbounded (even for
Hamiltonians with one-sided unbounded spectrum).} operator $S$\thinspace\ is
involutive on its domain. By an additional notational convention one gets this
relation into the form where it may be viewed as a special illustration of a
much more general operator algebra structure which is the famous
Tomita-Takesaki modular theory of operator algebras \cite{Bra-Ro}.

\begin{theorem}
(main theorem of the Tomita-Takesaki modular theory) Let ($\mathcal{A,H}%
,\Omega$) denote a weakly closed (von Neumann) operator algebra $\mathcal{A}$
acting on a Hilbert space $\mathcal{H,}$ with $\Omega\in\mathcal{H}$ a vector
on which $\mathcal{A}$ acts in a cyclic and separating manner. Then there
exists an antilinear closed involutive operator $S$ which has the dense
subspace $\mathcal{A}\Omega$ in its domain such that
\begin{align}
SA\Omega &  =A^{\ast}\Omega,\,A\in\mathcal{A}\\
S  &  =J\Delta^{\frac{1}{2}},\,J\Delta=\Delta^{-1}J\nonumber
\end{align}
The polar decomposition of $S$ leads to an antiunitary $J$ and a positive
$\Delta$ which in turn defines the unitary modular group $\Delta^{it}.$ Their
significance results from their adjoint action on the algebra
\begin{align}
J\mathcal{A}J  &  =\mathcal{A}^{\prime}\\
\sigma_{t}(A)  &  \equiv\Delta^{it}A\Delta^{-it}\in\mathcal{A}\nonumber
\end{align}
The modular automorphism $\sigma_{t}$ fulfills the following KMS property
(with $\beta=-1,$ see below$)$%
\begin{equation}
\omega(\sigma_{t}(A)B)=\omega(B\sigma_{t-i}(A)B),\,\omega(\cdot)\equiv\left(
\Omega,\cdot\Omega\right)
\end{equation}
and depends only on the state $\omega$ (and not on its implementing vector
$\Omega).$
\end{theorem}

The relation to the HHW work is most directly established via the validity of
the KMS property which replaces the Gibbs formula in the thermodynamic limit%

\begin{align}
&  \omega_{\beta}(\alpha_{t}(A)B)=\omega_{\beta}(B\alpha_{t+i\beta}(A))\\
\exists\,F_{A,B}(z)\,,\,F_{A,B}(t)  &  =\omega_{\beta}(B\alpha_{t}%
(A)),\,F_{A,B}(t+i\beta)=\omega_{\beta}(\alpha_{t}(A)B)\nonumber
\end{align}
where the second line expresses the analytic content of the KMS condition in
more careful terms: there exist an analytic functions $F_{A,B}(z)$ which
interpolates between the thermal expectation values of operator products taken
in different orders; this function is analytic in the strip $0<Imz<\beta$ and
has continuous boundary values on both margins which relate to the two
different orders.

The nontriviality of the T-T proof relates to the fact that the
J-transformation property into the commutant and the automorphic action of
$\sigma_{t}$ turns out to be much harder. Specializing again to the Gibbs
setting, its HHW tensorproduct structure re-appears in the modular setting in
the following way
\begin{align}
&  H_{\operatorname{mod}}\equiv\pi(H)-\pi_{r}(H)\label{mod}\\
&  \Delta^{it}\equiv e^{-i\beta tH_{\operatorname{mod}}},\,\,S=J\Delta
^{\frac{1}{2}}\nonumber\\
&  H_{\operatorname{mod}}\psi_{\kappa_{0}}=0,\,\,\Delta^{it}\psi_{\kappa_{0}%
}=\psi_{\kappa_{0}}\nonumber\\
&  \Delta^{-it}\pi(A)\Delta^{it}=\pi(\alpha_{\beta t}(A)),\,\,\alpha
_{t}(A)=Ade^{iHt}(A)\equiv e^{iHt}Ae^{-iHt}\nonumber
\end{align}
Whereas in vacuum QFT the energy operator $H$ (obtained by integrating the
energy density) is the generator of the translation, in the heat bath
situation it is the doubled Hamiltonian $H_{\operatorname{mod}}$ which leaves
the thermal reference state invariant, generates the translation symmetry and
has finite fluctuations in the thermodynamic limit. The doubling of Fock space
through tensoring can be used to arrange the computational scheme in such a
way that the recipe parallels the Feynman rules for the zero temperature case.
In this form it gained widespread popularity under the name ``Thermo Field
Theory''\footnote{The authors who introduced it \cite{Umezawa} appearantly
were not aware of the close connection to the older Haag-Hugenholz-Winnink
KMS-based formulation and this had the effect that the majority of thermal
practitioners up to this date remained unaware of the connection with the more
fundamental modular theory (see however \cite{Ojima})} (at this School it was
used in M. C. Abdalla's talk). Besides of being more general, the KMS setting
is more faithful to the main aim of theoretical physics which is the
de-mystification of nature. According to the above remarks the tensor
structure of the Thermo Field Theory is lost in the thermodynamic limit in
which case one can simply use the modular $J$-operation to define the
``Tilde'' fields of TFT \cite{Ojima}.

The KMS framework is also very successful in showing the equivalence between
the Matsubara imaginary time (discrete energy) and real time (e.g. TFT)
formulations. The mathematical proof is bases on the use of very nontrivial
Carlsonian type of theorems \cite{Viano}. It also leads to an extension of the
KMS analyticity region (the relativistic KMS \cite{B-B1}) and to an
understanding of the perseverance of dissipative effects in the timelike
asymptotic behavior \cite{B-B2} which is important for the avoidance of
perturbative infrared divergencies.

In the next section it will be shown that the modular framework is capable of
incorporating both the heat bath- and the localization- caused thermal properties.

\section{Thermal aspects caused by vacuum polarization on horizons}

The Reeh-Schlieder theorem \cite{Haag} of QFT (the localization-generated
``operator-state'' relation in the more folkloristic terminology often used in
conformal QFT) insures that these properties are always fulfilled as long as
the localization region has a nontrivial causal disjoint. Although the
Tomita-Takesaki theorem asserts that the modular KMS automorphism always
exists in these cases, the physical interpretation up to now has been
restricted to cases of geometric (diffeomorphism, non-fuzzy) action of the
modular group\footnote{The vacuum modular group of a double cone algebra is an
example of what is meant by fuzzy action. It is believed that in the standard
formulation (where such algebra is generated by smeared fields with double
cone supported test functions) the fuzzy modular actions are
support-preserving maps of test function spaces (probably related to
pseudo-differential operators) which are asymptotically geometric at the
causal horizon \cite{S-W}.} $\sigma_{t}$ which in the context of Minkowski
space leaves only the Lorentz boosts of wedge region (whereas in CST there are
many models with horizon-preserving Killing symmetries). The best known
illustration without curvature is Unruh's Gedankenexperiment in which the
observables localized in a Rindler wedge region bounded by a causal horizon
are realized by a family of uniformly accelerated observers whose Hamiltonian
is proportional to the Lorentz boost. The relation with the Tomita-Takesaki
modular theory was first noticed by Sewell who observed on the basis of prior
work on the modular theory of wedge algebras by Bisognano and Wichmann that
Unruh's Gedankenexperiment can be viewed as a physical realization of the B-W
mathematical results (for a simple but enlightening presentation see
\cite{Sewell}).

This raises the question whether the Bekenstein area behavior could also be
seen as a (classical) manifestation of the same vastly general local quantum
physics mechanism which is responsible for the appearance of a temperature via
vacuum polarization from quantum localization. Trying to answer this question
with standard box quantization methods and ad hoc cut-offs (in order to obtain
an entropy via degree of freedom counting) proved to be inconclusive
\cite{Sorkin}. According to the above ideas the relevant question should be
whether by physically motivated ideas (i.e. by remaining within the given
local theory \ and thus avoiding ad hoc locality-violating cut-offs) one can
associate a localization entropy with the Lorentz boost in its role as the
modular group of the wedge algebra. If one would be able to show that in this
Unruh test case there exists an area density of localization entropy which is
the counterpart of the well established horizon-affiliated KMS localization
temperature, then the present thesis that the existing successful framework of
QFT, if extended by some new concepts derived form the old principles, would
gain strength and the many attempts to invoke speculative physics and the blue
yonder (borrowing a phrase from Feynman) may use their strong spell which they
exerts especially on young members of the physics community. If this (despite
the encouraging signs below) should turn out to be disproved by future more
detailed computations, one at least would have a theoretically more solid
point of departure and a better guide for speculative endeavors.

There is obviously no chance in QFT to directly assign an entropy to the
modular operator of the Unruh wedge situation (which is the Lorentz boost).
Although the lightfront holography of the Bisognano-Wichmann/Unruh-Rindler
wedge algebra is easily seen to map the boost into the generator of scale
transformations, one is still stuck with a non trace class operator. The
restriction of the global vacuum to the horizon algebra assures the
thermalization in the sense of a geometrically determined (Hawking)
temperature, but it does not help in getting closer to trace class properties,
although (as a result of the loss of transverse vacuum polarization, see
below) it reduces the problem of understanding of an area density to that of
entropy of the vacuum restricted to a of a chiral algebra  restriction to a
halfline algebra (=halfcircle in the compact description of chiral theories).
The essential step for getting a density matrix from the local restriction of
the vacuum is to allow the halfline  localization to be ``fuzzy'' by an
``$\varepsilon$-roughening'' of the boundary endpoints. This \textit{split}
process of leaving a distance $\varepsilon$ between the halfcircle
localization of the chiral algebra and that of its commutant (the opposite
halfcircle) is the opposite analog of the thermodynamic limit namely instead
of starting from Gibbs states in order to approach the KMS thermodynamic limit
state one wants to search for density matrix states associated with fuzzy
boundaries which in the limit $\varepsilon\rightarrow0$ lead back to the KMS
state. According to our previous considerations this split-restricted vacuum
state is a vector in the two-fold tensor product of a (ground state) Fock
space with itself. As in the Thermo Field formalism this vector is highly
entangled and becomes impure upon restriction to the ``physical'' tensor
factor\footnote{In TFT the physical algebras are generated from the original
observables and the commutant (the ``tilde'' operators) do not occupy any
``geometric territory'', whereas in the horizon-caused thermalization the
commutant is localized ``behind the horizon''. There is however the curious
observation that for systems with no transverse direction in their holographic
projections (i.e. 2-dimensional QFT), the distinction between the heat bath
``shadow world'' and the real world ``behind the horizon'' becomes blurred
\cite{S-W}\cite{B-Y}.}. In the limit the thermalized physical vacuum becomes
orthogonal to the split tensorproduct vacuum, in fact both vacua are cyclic
and separating reference states which belong to inequivalent representation of
a suitably defined $C^{\ast}$ tensorproduct algebra \cite{Wald}. The simple
structure of local chiral algebras (generalized W-algebras) permits to argue
that the speed of vanishing of overlaps is dominated by powers of
$\varepsilon$ (whereas area densities of partition functions diverge according
to inverse $\varepsilon$-powers which suggests that the divergence of the
split entropy should go universally like -$\ln\varepsilon.$ Formulating this
expectation as an universality conjecture \cite{S3} one obtains the statement
that the holographic universality classes (different ambient matter content
may be holographically mapped into the same chiral theory) lead to
(class-dependent) numerical coefficients multiplying -$\ln\varepsilon$ so that
the split property can only determine finite ratios of area densities. Thus
the area density resulting from the split property can only be a
\textit{relative} entropy density; the holographic formalism together with the
split property can never produce a normalized entropy density. In fact in view
of all the black hole analogs one does not even want a normalized entropy on
this level of discussion because besides the principles of local quantum
physics (causal propagation in a local quantum context) we have not used
properties which would distinguish between the different analogs and the
Hawking gravitational black hole which would set the different scales in a
Bekenstein entropy argument for those analogs; one expects the surface
``gravity'' (the Unruh acceleration) related to the numerical factors between
the geometrically determined modular automorphism and the ``Hamiltonian'' to
set this scale. The absolute normalization can only come (as in the
Bekenstein-Hawking case) through the validity of a second fundamental
thermodynamic law in which the entropy is related to other quantities.
Bekenstein takes the classical Gibbs form of this law, but the problem in the
present setting would be to find out if and how such a law caqn be derived
outside the classical heat bath setting. Here we are entering an unknown area
of QFT in which however the questions seem to be well-posed.

From a pragmatic viewpoint the different steps in the argument all serve to
extract a well-defined additive (under correlation-free subdivisions) measure
of impurity for the horizon-restricted vacuum (alias wedge-restricted vacuum).
This is quite different from \cite{Sorkin}; the box of those authors should be
causally completed to a double cone, but even then a treatment paralleling the
present would be much more difficult as a result of the nongeometric nature of
the associated modular group \cite{S1}\cite{S2}.

The remainder of this section will be used to present the argument about the
absence of transverse vacuum correlations in the holographic lightfront
projection. For brevity (and pedagogical reasons) we limit the presentation to
the holographic lightfront projection of scalar free fields. In that case one
finds that in terms of Weyl generators the result looks as follows%

\begin{align}
&  W(f):=e^{iA(f)},\,\,A(f)=\int A(x)f(x)d^{4}x\\
&  W(g,f_{\perp})\longrightarrow W_{LF}(g,f_{\perp})=e^{iA_{LF}(g,f_{\perp}%
)}\,\nonumber\\
&  with\,\,\,A_{LF}(g,f_{\perp})=\int a^{\ast}(p_{-},p_{\perp})g(p_{-}%
)f_{\perp}(x_{\perp})\frac{dp_{-}}{2\left|  p_{-}\right|  }d^{2}p_{\perp
}+h.c.\,\\
&  \curvearrowright\left\langle W(g,f_{\perp})W(g^{\prime},f_{\perp}^{\prime
})\right\rangle =\left\langle W(g,f_{\perp})\right\rangle \left\langle
W(g^{\prime},f_{\perp}^{\prime})\right\rangle \text{ }if\,\,suppf\cap
suppf^{\prime}=\emptyset\nonumber
\end{align}
The second line formulates lightfront restriction on the dense set of wedge
supported test functions which factorize into a longitudinal and a transverse
part \cite{S1}\cite{S2}. The third line is the statement that holographically
projected Weyl generators (and therefore also the algebras they generate) have
no transverse fluctuations; the holographic projection compresses all vacuum
fluctuations into the lightlike direction. This reduces the problem of
horizon-associated entropy to the problem of looking for an area (the area of
the edge of the wedge which limits the upper horizon) density of entropy as
mentioned before. and should be interpreted as the localization entropy of a
halfline in a chiral theory. It turns out that the lightfront holography leads
to a QFT with a seven parametric symmetry subgroup of the Poincar\'{e} group
which contains in particular a transverse Galilei group which results from the
holographic projection of the ``translations'' contained in Wigner's
3-parametric ``little group'' of the lightray in the lightfront. This is also
true in the general non-free situation.

We will skip the generalization of the proof to interacting theories since an
explanation of the methods (involving modular inclusions and intersections
\cite{S1}\cite{S2}) goes beyond the scope of this talk.

\section{Concluding remarks}

The main aim of these notes was two-fold, on the one hand we have recalled
that there exists a unified formalism for heat bath and localization caused
thermalization and on the other hand we emphasized that the most startling
difference between the two cases shows up in the total absence of vacuum
transverse polarization which is the prerequisite for area proportionality of
entropy. These considerations did not yet solve the existence of a
horizon-associated (relative) area density of entropy. In order to arrive at a
Bekenstein like formula one still has to prove a universality conjecture and
(for its normalization) and derive a second thermodynamic law in which this
quantum localization entropy enters. However the remarkable area dependence of
any would-be entropy (versus the standard volume dependence of heat bath
entropy) is already secured on the present level of understanding. The area
density of entropy (assuming that the conjectures can be established) inherits
the universality of the holographic projection\footnote{The holographic
universality classes are in some sense bigger than the short distance classes
since the resulting chiral algebras are one-dimensional and local i.e. there
can be no anomalous dimensions (non halfinteger critical indices).}. However
(in line with the experience that classical laws suffer modifications which
depend on the kind of quantum matter) in the presence of quantum matter one
perhaps should not expect total universality of the area density as in
Bekenstein's formula.

\end{document}